# The role of scientific output in public debates in times of crisis: A case study of the reopening of schools during the COVID-19 pandemic

Gabriela F. Nane[1], François van Schalkwyk[2], Jonathan Dudek[1,3], Daniel Torres-Salinas[4], Rodrigo Costas[2,3] and Nicolas Robinson-Garcia[1]

[1]Delft Institute of Applied Mathematics, Delft University of Technology, Delft, The Netherlands
[2]DST-NRF Centre of Excellence in Scientometrics and Science, Technology and Innovation Policy, Centre for Research on Evaluation, Science and Technology, Stellenbosch University, Stellenbosch, South Africa
[3]Centre for Science and Technology Studies, Leiden University, Leiden, The Netherlands
[4]Departamento de Información y Comunicación, Universidad de Granada, Granada, Spain

## Abstract

In exceptional circumstances such as health pandemics, it is to be expected that policy actions are supported by a balanced use of scientific information to support decision-making that impacts the lives of citizens. However, situations in which no scientific consensus has been reached due to either insufficient, inconclusive or contradicting findings place strain on governments and public organizations which are forced to take action under circumstances of uncertainty. In this chapter, we focus on the case of COVID-19, its effects on children and the public debate around the reopening of schools. The aim is to better understand the relationship between policy interventions in the face of an uncertain and rapidly changing knowledge landscape and the subsequent use of scientific information in public debates related to the policy interventions. Our approach is to combine scientific information from journal articles and preprints with their appearance in the popular media, including social media. First, we provide a picture of the different scientific areas and approaches, by which the effects of COVID-19 on children are being studied (e.g., transmission, infection, severity, etc.). This provides a snapshot of the scientific focus and priorities in relation to COVID-19 and children. Second, we identify news media and social media attention around the COVID-19 scientific output related to children and schools. We focus on policies and media responses in three countries: Spain, South Africa and the Netherlands. These countries have followed very different policy actions with regard to the reopening of schools and represent very different policy approaches to the same problem. We analyse the activity in (social) media around the debate between COVID-19, children and school closures by focusing on the use of references to scientific information in the debate. Finally, we analyse the dominant topics that emerge in the news outlets and the online debates. We draw attention to illustrative cases of miscommunication related to scientific output and conclude the chapter by discussing how information from scientific publication, the media and policy actions shape the public discussion in the context of a global health pandemic.

## Introduction

The COVID-19 pandemic has turned the world upside down. While it rapidly became clear that certain population groups are more at risk, with the elderly and adults who have underlying health conditions being at higher risk of developing severe illness from COVID-19, much uncertainty remained surrounding children (Rajmil, 2020). The uncertainty characterized both the infection rates among children as well as their range of symptoms. As empirical evidence accumulated at tremendous pace, fewer infection rates have consistently been reported in children compared with adults, as have the milder symptoms (Cruz and Zeichner 2020, Götzinger et al. 2020, Goldstein et al. 2020). Very few cases in children have been linked to severe symptoms, such as multisystem inflammatory syndrome and Kawasaki-like disease(Viner and Whittaker 2020, Webb et al. 2020).

Nonetheless, the infection rate in children is considered to be biased, given the testing policies in many countries. For example, in the Netherlands only those (mildly) symptomatic are eligible for COVID-19 testing. And whereas everyone with cold-like symptoms could be tested over the summer, from September 26th 2020 up until the time of writing (December 2020), children of 12 years old or younger have been placed under special testing rules, and were only allowed to be tested following the presentation of serious symptoms. Adults and children older than 13 years of age could still be tested if they presented cold-like symptoms. The special testing rules stipulated being really sick or being in contact with someone who had tested positive for COVID-19. Children with cold-like symptoms have been allowed to attend school since September 26th 2020.

A substantial body of work about COVID-19 and children has focused on the role of children in spreading the virus. The role of children in the transmission of the virus was questionable from the onset of the pandemic, and the topic is still under debate. To illustrate: "Children may play a major role in community-based viral transmission" according to Cruz and Zeichner (2020) whereas Ludvigsson (2020) reports that "children are unlikely to spread the coronavirus". Media reporting reflects this debate. A recent news article in *Nature* (Lewis, 2020) reports that young children are unlikely to spread the virus, whereas an article in *The Conversation* (Hyde, 2020) states that "children may transmit coronavirus at the same rate as adults".

Despite the ongoing debate about the role of transmission in children, schools received distinct attention early on. The enclosed and crowded environment, prone to poor ventilation and where children and educators spend 6 to 8 hours daily, creates the conditions for a high-risk environment. Hence, strict measures have been necessary in the face of scientific uncertainty and closing schools was among the first measures taken worldwide to reduce the spread of the virus. China and Mongolia were the first countries to close schools in the middle of February 2020, followed by some schools in Italy and San Marino at the end of February 2020. By March 31st, schools in 193 countries were closed due to COVID-19.

The reopening of schools has been part of the first steps in the easing of lockdown restrictions. Whereas 43 countries reopened schools partially and other 40 countries fully reopened schools by June 15th, many other countries in the northern hemisphere postponed the reopening of schools until after the summer holidays. In a considerable number of countries, school reopening has been perceived as being safe, given the low levels of community spreading. Nonetheless, when community spread levels were not low, hot spots of infections were reported at a school and at a school camp (Stein-Zamir et al., 2020; Torres et al., 2020; Szablewski et al., 2020). An ECDC report on COVID-19 in children and the role of school settings in COVID-19 transmission concludes that "there is limited evidence that schools are driving transmission of COVID-19 within the community, however there are indications that community transmission is imported into or reflected in the school setting". Macartney (2020) has reported low transmission rates in the New South Wales educational system during the first wave of the COVID-19 pandemic. Even though most schools remained open in Australia during the first wave, class sizes were reduced during the peak. For some parts of the country (e.g. Melbourne) with high community viral spreading, schools did close.

There is thus no scientific consensus reached so far from the empirical research on children's role in the transmission of the coronavirus. Nonetheless, decision-making concerning school reopenings and closure could not and cannot wait for greater scientific consensus.

This chapter presents an exploratory attempt at tracing social discussion around a scientific topic under debate in the midst of a global pandemic, combining quantitative and qualitative methods. We focus on the case of COVID-19 and its effects on children which inform the public debate around the reopening of schools. We do so to better understand the relationship between policy interventions during an uncertain and rapidly changing knowledge landscape and the subsequent use of scientific information in public debates related to the policy intervention during a crisis. Our approach is to combine scientific information with their appearance in the popular media, including social media.

We investigate the reopening of schools in three different countries (Spain, South Africa and the Netherlands), each of which introduced different policy measures, with the aim of analysing the societal reception of scientific findings in three different national and political contexts. In the case of the Netherlands, after an initial lockdown, schools reopened in May, at quite an early stage of the first wave of the outbreak and have been opened ever since (with the exception of summer holidays). In South Africa, the outbreak took place in March and schools closed until early June, just to close again a month later due to the rise of infections, and reopen again in August. Finally, Spain has been one of the European countries with the most restrictive measures at the early stage of the pandemic. Schools did not open until after the summer holidays. For each of these countries, we retrieved information related to these policy interventions as well as the dates on which announcements were made.

## Conceptual framework

Our study is situated at the intersection between science, politics and society. We take as our starting point the fact that public communication about science is inherently political, and adopt Scheufele's (2014) conceptualisation of science communication as political communication. Such a conceptualisation takes into account the broader political contexts in which science–public interactions occur, how stakeholders compete for attention in the political sphere, and how publics interact with the scientific information they encounter in the media — information which may often be contradictory as well as overwhelming in complexity and volume. Moreover, the information is usually presented via multiple traditional and online channels, and may change rapidly during times of heightened uncertainty such as the COVID-19 pandemic.

The COVID-19 pandemic presents a unique case of science communication as political communication because the threat posed by the virus is both immediate and global. The consequence is a simultaneous and rapid response to the pandemic by scientists, politicians and the public alike.

Science responds to the crisis by conducting research aimed at understanding the behaviour of the virus and to develop effective responses to containing its spread. To share new truth claims with other scientists with the objective of accelerating the discovery of effective responses to the pandemic, findings from scientific research on the novel coronavirus are fast-tracked for publication in preprints and scientific journals. These findings are also communicated to policy-makers and to the public, either indirectly via the media or directly via briefings, press releases and social media.

While science advances understanding of the coronavirus, political decisions are taken to control the spread of the virus in order to protect society. These political decisions constitute policy moments in response to the pandemic. Political decisions are informed by the local context, including the progression of the outbreak and the prevailing political climate, and will to varying degrees be influenced by the available science. The degree of influence that science exerts over political decision-making is not only likely to depend on context but is likely to change over time as the perception of the threat changes and as other political issues such as the socio-economic impact and the infringement on citizens' constitutional rights begin to challenge the legitimacy of the measures taken to control the pandemic.

The social response to policy moments during the pandemic is reflected both in the mainstream and in the social media as citizens process and debate both the available scientific information and the political decisions taken. At the same time, scientists themselves seek to popularise or make more accessible the latest scientific information about the virus. Scientists are also attentive to the media leading to the 'medialization' of science, that is, to increasing links and interplay between science and the media (Weingart et al., 2012).

## Data and methods

Our point of departure is the scientific output generated around the COVID-19 pandemic as it relates to children. This corpus of literature includes scientific papers with a broad scope of topic, including the mental and social effects related to policy interventions, effects of the lockdowns, the closure of schools, and medical issues related with the infection, transmission and diagnosis of COVID-19 in children. From this set of publications, we trace signals of discussion in social media and news media platforms, as a means to establish a link between the scientific and societal realms.

## Data collection

The data collected for this study was extracted from a variety of sources: scientific publications, news outlets, and social media discussions and policy interventions. Since the outbreak of the pandemic, different community- and organization-led initiatives have been conducted to make scientific publications on COVID-19 openly accessible. In this study, we made use of the COVID-19 Open Research Dataset (CORD-19) and the World Health Organization (WHO) COVID-19 Global literature on coronavirus disease database. These two databases are of special interest due to the combination of sources they include, containing not only studies published in scientific journals but also preprints from the main global repositories (e.g., BioRxiv, MedRxiv, SSRN, etc.).

We downloaded the two complete databases on October 15, 2020. Table 1 shows some descriptive values of the size of the database at the time. We searched within the title and abstract fields for documents containing the words ‘children’ and ‘schools’. After merging the ‘Pubs children’ documents of both datasets, a total of 5,713 publications were retrieved. This is our final set of scientific publications from which we trace their (social) media reception.

**Table 1. Descriptive values of the publication databases used in the study.**

| database | Pubs | Pubs in= 2020 | % DOI= in 2020 | Pubs children | % DOI children |
|---|---|---|---|---|---|
| WHO | 113,105 | 103,084 | 26.67 | 4,434 | 30.42 |
| CORD-19 | 314,001 | 220,251 | 50.87 | 9,380 | 52.03 |
| Merged final set (via DOI) | -- | -- | -- | 5,713 | 100 |

Note: Number of total publications by database [Pubs], publications in 2020 [Pubs in 2020], share of publications with Document Object Identifier (DOI) [%DOI in 20202], number of publications related to children and schools [Pubs children] and share of publications with a DOI related to children and schools [%DOI children]

The two databases (WHO and CORD-19) do not represent distinctive sets of publications, having quite substantial overlap. In order to avoid duplicates, the two databases were merged and cleaned. For a reliable merging of the two databases, as well as for the further tracing of the (social) media reception of the publications, it was necessary to count with unique document identifiers (e.g., PubMed Identifiers, Digital Object Identifiers, etc.). Particularly Digital Object Identifiers (DOI) are commonly assigned to scientific publications to univocally identify scientific documents across databases and the web-at-large. The main inconvenience of using DOIs is that we can only identify and combine publication data for half of the papers included in the CORD-19 database and a third of those included in the WHO database (Figure 1).

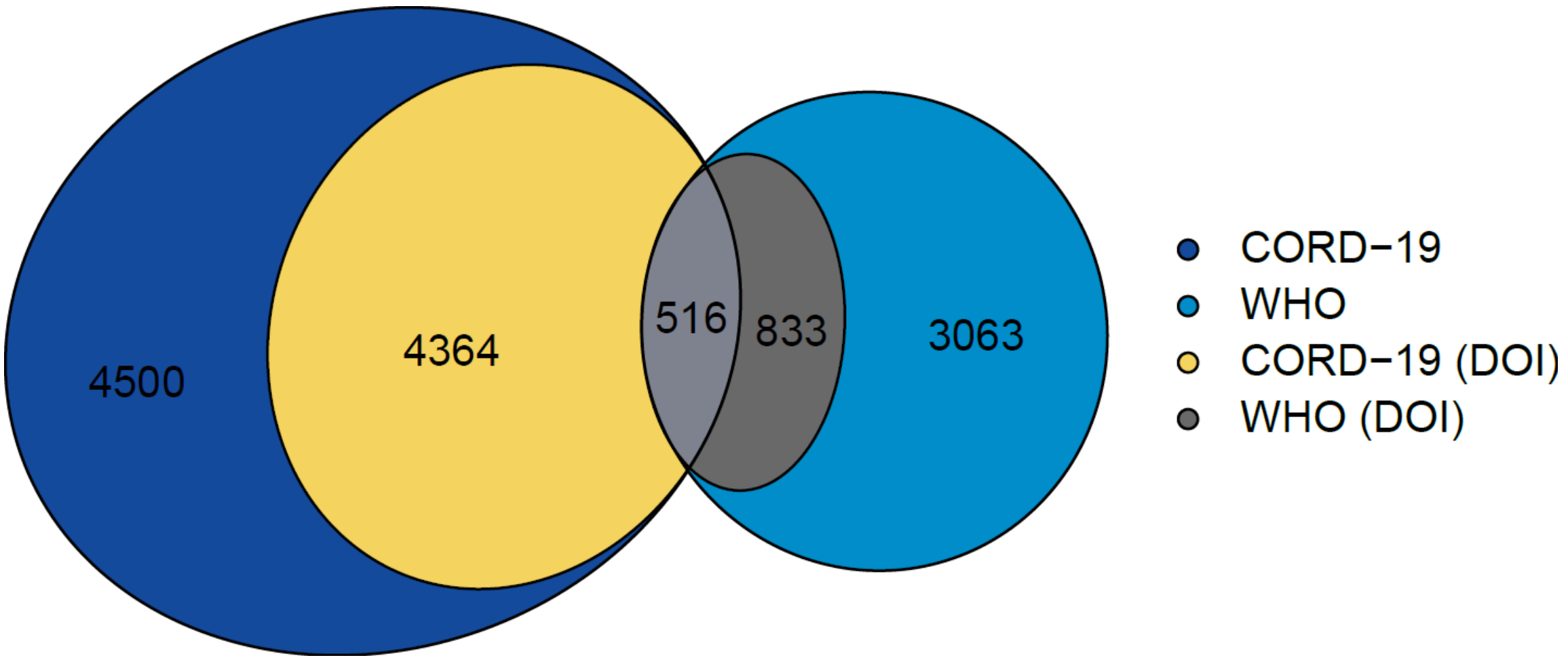

**Figure 1. Number of publications from 2020 related to children and schools, indexed in the CORD-19 and WHO databases.**

A final number of 5,713 publications along with their DOIs have been collected in our final dataset of scientific output. We proceeded to identify news outlets and social media discussions around the scientific publications in our dataset. News media items mentioning a DOI in our set were identified with data from Altmetric.com, retrieved in October 2020. From a total of 19,922 news items found globally for the set of DOIs in our database, 424 news articles could be identified as originating from the Netherlands, Spain, or South Africa. This was done by matching the URLs of the news outlet coming from Altmetric.com with the URLs of Dutch, Spanish, and South African national newspapers and broadcasting services, as extracted from Wikipedia and other websites listing news outlets. The final list of news outlets from each country was verified and curated manually. We identified 200 news items from Spain, which referenced 81 distinct DOIs. In South Africa, 79 news pieces referenced 72 distinct DOIs and in the Netherlands, 145 news items referenced 83 distinct DOIs. The titles and short abstracts of the news articles (where available in the data from Altmetric.com) were analyzed manually for our study.

Twitter data on mentions of publications were also collected from Altmetric.com in October 2020. This included any tweet identified by Altmetric.com that refers to a DOI in our set of publications. More detailed Twitter data was rehydrated directly from Twitter (using the Twitter API) on December 2nd, 2020. This resulted in a total number of 540,615 tweets, covering 66.7% (3,811) of the publications. The first identified tweet was on January 14, 2019, and the last recorded one on October 24, 2020. From our tweet data, 65.8% (182,548) of the 277,419 distinct Twitter users provided geolocation information. This allowed us to link tweets to the three countries selected in our study. We identified 16,548 tweets with a Spanish geolocation, which referenced 932 distinct DOIs. Much less Twitter activity was captured in the cases of the Netherlands and South Africa. In the Netherlands, 1,478 tweets were collected, linking to 229 distinct DOIs, whereas in South Africa, 1,062 tweets could be linked to 290 distinct DOIs.

Lastly, data on policy interventions regarding the closure and reopening of schools was retrieved from the UNESCO Institute for Statistics, which includes daily global information on the state of schools since the outbreak of the pandemic. With regard to the announcements and specificities of the measures, we manually searched national news media platforms.

## Semantic analysis

We explored the general semantic configuration of the publications selected by means of co-word maps, extracted from the titles of the articles identified. We employed natural language processing tools to extract noun phrases from titles. We then created a binary co-occurrence matrix to build and visualize the final network. This network provided a baseline to visually inspect the whole body of literature related with COVID-19 and children. Moreover, it also offered a baseline on which to overlay Twitter and news media activity. This was done by coloring the nodes of the network (noun phrases) based on the number of mentions received by the papers containing such terms. Thus, terms with a higher intensity in the color grading belong to papers that are highly tweeted by users of a given country. Figure 2 provides a visual aid to help the reader better understand and interpret the contents of such maps.

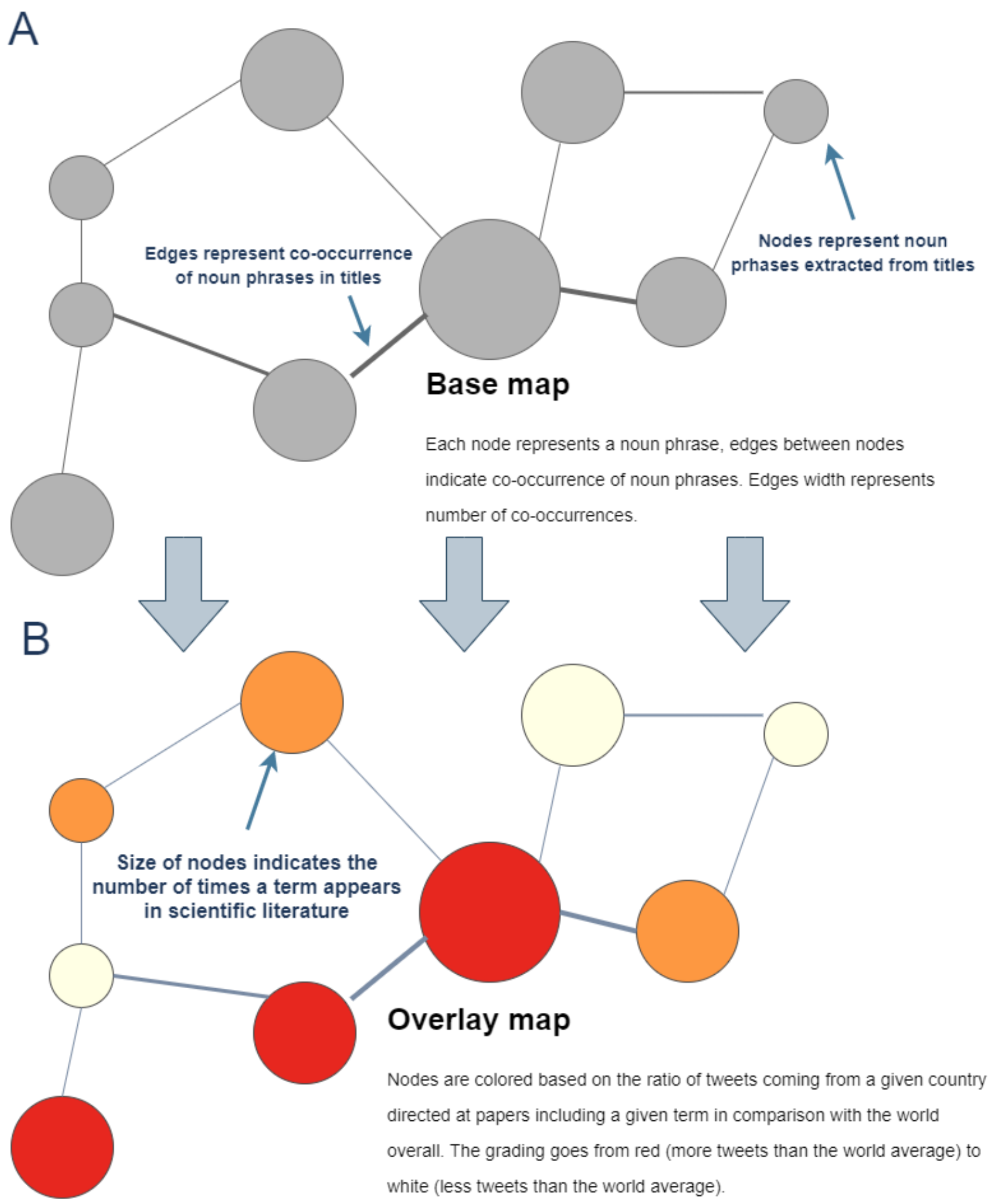

**Figure 2. Graphical representation explaining overlay maps and how to interpret them correctly. Graph A shows the base map constructed with the complete set of scientific literature, while Graph B overlays tweet mentions to publications by coloring nodes based on the intensity of the tweets.**

The data collected on news outlets and tweets were used to investigate the activity and topics covered in the (social) media when reporting scientific outputs, for each country. Overlay maps were used to show those topics. We include overlay maps for tweets in this chapter. We mention that single noun phrases "covid" and "children" were removed from the maps since they are redundant for our analysis. The overlay maps for news outlets are uploaded on figshare. For illustrative purposes, we selected examples to study differences in the reporting of scientific literature in the news media, by individually reading and analyzing the contents of selected scientific and news articles, and the tweets text. The specific findings will be reported for each country in the following section.

# Results

## Semantic structure of the research on children, schools and COVID-19

From a literature review and a clustering analysis, it became apparent that the scientific output consisting of 5,713 publications since the start of 2020 focused on the risk of infection, as well as on the development of mild or severe illnesses from COVID-19 in paediatric cases. The large number of asymptomatic cases, as well as the role of children in spreading the virus also received considerable attention. Despite the main focus on the physical health of children during this pandemic, attention was given to the psychological effects of quarantine and lockdown during COVID-19 (Orgilés et al., 2020; Idoiaga et al., 2020). The vulnerability of children during this pandemic was also researched (Haffejee and Levine, 2020; Fouche et al., 2020). Finally, inequality in home-schooling during the pandemic also received attention from scientists (Bol, 2020).

Reopening schools does not only involve the health and well-being of children but also of adults who come in close contact with children. In this regard, research focused on the risk of severe COVID-19 illness among teachers and among adults living with school-aged children (Gaffney et al., 2020).

Broad themes such as *infection*, *development of severe symptoms*, *transmission*, and *social and psychological impacts of school closures* were identified as dominant in the scientific literature on the subject of children and COVID-19. The underlying map is available on Figshare.

A manual search identified research output on children's health and school reopening within the three countries. The focus on the scientific output in the three countries varies. In Spain, significant attention has been given to hospitalization and to severe cases of children with COVID-19. Tagarro et al. (2020) reported on the early screening and severity of coronavirus in Madrid by investigating data from 365 tested children in the first two weeks of March 2020. Attention has also been given in Spain to the psychological effects and well-being of children (Orgilés et al., 2020; Idoiaga et al., 2020).

In the Netherlands, research has focused on the transmission of the virus by children. A study on 54 households "suggest lower point estimates for transmissibility of infection to close contacts from children aged under 19y, and higher point estimates for adults aged over 70y when compared to persons aged 19-69y" (RIVM, 2020). Alsem et al. (2020) reported on the effects of the pandemic on paediatric rehabilitation, whereas Bol (2020) investigated the inequalities in home-schooling.

In South Africa, research has been conducted on child protection and resilience (Fouche et al., 2020; Haffejee and Levine, 2020) and well-being (van Bruwaene et al., 2020), indicating a focus on the social aspects of school closures and lockdown. Additionally, attention has been given to multisystem inflammatory syndrome in children in South Africa (Webb et al., 2020).

The scientific output results in mixed evidence of infection and transmission as they pertain to children. The limitations of the scientific studies and the consequent levels of uncertainty were conveyed when reporting findings. This is, however, not a unanimous approach. For example, a viewpoint in the *Archives of Disease in Childhood* (Munro and Faust, 2020), is entitled "Children are not COVID-19 super spreaders: time to go back to school". The title appears to be inflated by the urgent need for policy decisions. The authors write "At the current time, children do not appear to be super spreaders. Sero-surveillance data will not be available to confirm or refute these findings prior to the urgent policy decisions that need to be taken in the next few weeks such as how and when to reopen schools." They continue "Governments worldwide should allow all children back to school regardless of comorbidities. Detailed surveillance will be needed to confirm the safety of this approach, despite recent analysis demonstrating the ineffectiveness of school closures in the recent past (Viner et al. 2020b). The media highlight of a possible rare new Kawasaki-like vasculitis that may or may not be due to SARS-CoV2 does not change the fact that severe COVID-19 is as rare as many other serious infection syndromes in children that do not cause schools to be closed". The title suggests no uncertainty regarding the role of children in the transmission of coronavirus. A more uncertain approach is taken on the severity of symptoms, where a rare disease "may or may not be" attributed to the novel virus.

## Analysis per country of policy, news outlets and social media response

We investigated the (social) media response to the policy actions related to school closures and reopening. For this, we monitored news outlets and Twitter activity and investigated the extent to which they overlapped temporally with the policy actions. Moreover, we analyzed the topics covered by tweets and in news articles using the overlay maps. For exploratory purposes, we manually inspected the titles of news articles, as well as the full text of selected news items, and the content of selected publications. The analysis is presented for each of the three countries in our study.

### Spain

While in the Netherlands and South Africa schools reopened after around two months of closure, in Spain school reopening was delayed until after the summer holidays. Figure 3 depicts the announced and implemented measures, in chronological order, both at the national, as well as the regional levels. The policy measures registered no difference between primary and secondary schools. The figure also includes the timeline distribution of the news outlets and tweets in our database, which have been identified as originating from Spain. A total of 188 news articles and 15,603 tweets were identified between the beginning of February and the end of September 2020. News articles on the topic registered brief appearances before the school closure in March, as well as more consistent appearances around the reopening of schools in September. As for tweets, we can observe small peaks around the time of the announcements in March, as well as shortly before and after the schools reopening in September. Further activity has been registered during the school closure, with peaks around end of April, when the government announced a plan for easing lockdown restrictions, as well as in July and August, when no other policy intervention has been announced nor occurred.

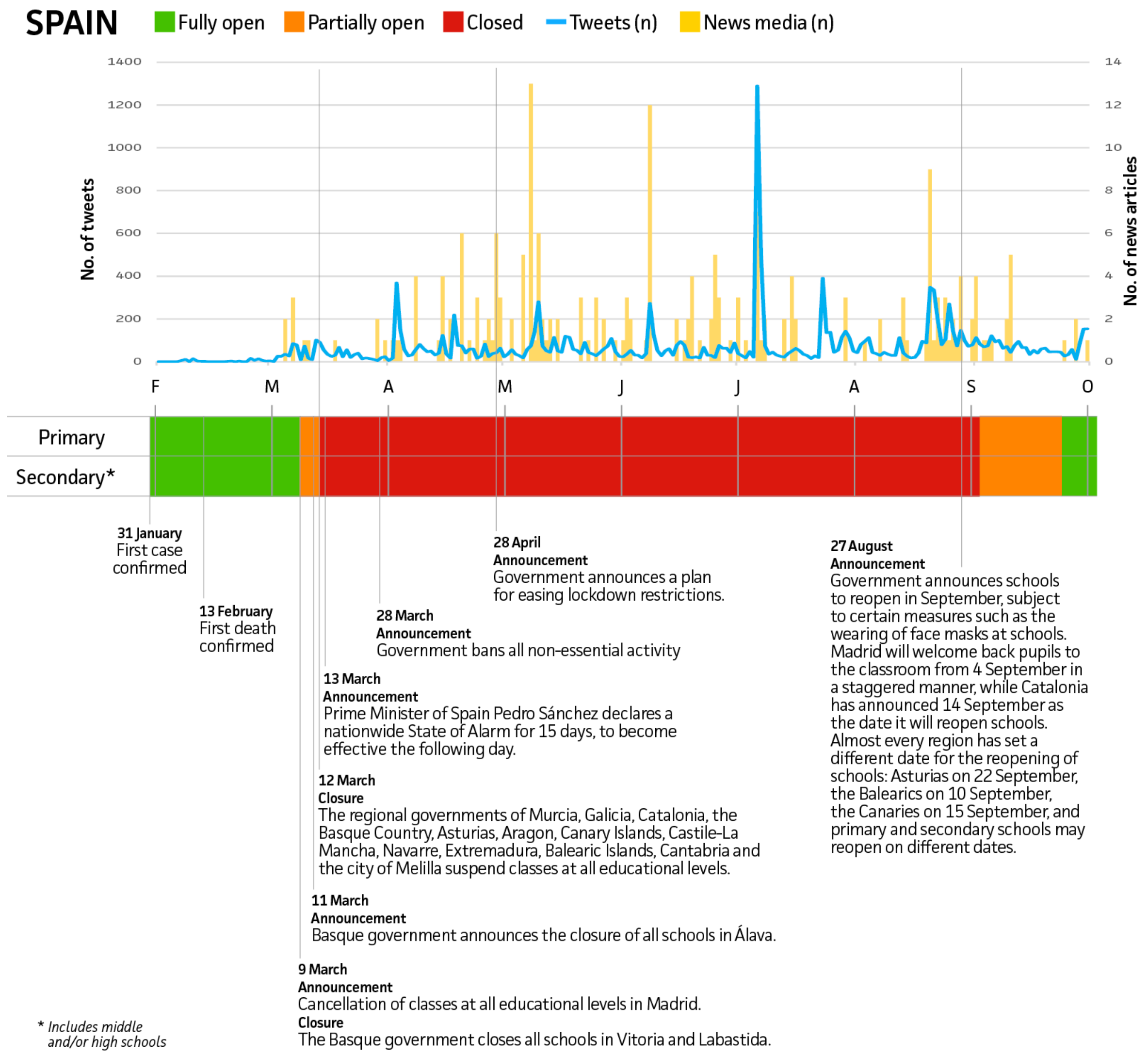

**Figure 3. Timeline of announcements and implementation of school closure and reopening in Spain, along with the distribution of tweets (on the left y-axis) and news items (on the right y-axis) mentioning scientific articles.**

The highest number of tweets in early July is the result of mentions received by a nationwide, population-based seroepidemiological study (Pollán et al., 2020) on the prevalence of SARS-CoV-2 in Spain (ENE-COVID). A total of 1,906 tweets followed soon after the article was published in the *Lancet*, at the beginning of July. The second-highest tweeted article (739 tweets) reports on paediatric severe acute respiratory syndrome (Yonker et al., 2020) and received distinct attention before the school reopening, as well as in September. Similarly, attention was paid before the school reopening to the safety of reopening (primary) schools during pandemic (Levinson, et al., 2020; Mallapaty, 2020).

In the case of news mentions, we do not observe the same activity pattern. There are almost no mentions in March, with sustained but low activity between April and July, and recurrent peaks on specific days at the beginning of May, June and July. This is followed by more constant activity in the media at the end of August, prior to the reopening of schools. The most mentioned article (20 mentions) received most of the attention in April. The article presents findings on the potential impact of the summer season on slowing the pandemic (Jüni et al., 2020) weighed against an alternative hypothesis that school closures account for such slowing. The two next most mentioned papers have 17 mentions each. In one case, Pollán et al. (2020) discuss the results of a nationwide screening undertaken in Spain between April and May. This paper concludes that at the time of the survey, there was a seroprevalence of around 5% with lower figures for children (<3.1%) and a third of the positive cases being asymptomatic. The third study focuses specifically on children tested positive (Yonker et al., 2020) aiming at “understanding the potential role children play” (p. 45) in the pandemic.

An overview of the topics covered by tweets from Spain is depicted by a VOSviewer map in Figure 4. The nodes in the map present the co-occurrence of the most relevant keywords identified from the titles of the 5,713 articles in our dataset. The color coding of the map reflects the prevalence of mentions of those articles in tweets from Spain relative to the worldwide collected tweets: The darker the color, the more focus on the keywords relative to the worldwide tweets on the topic.

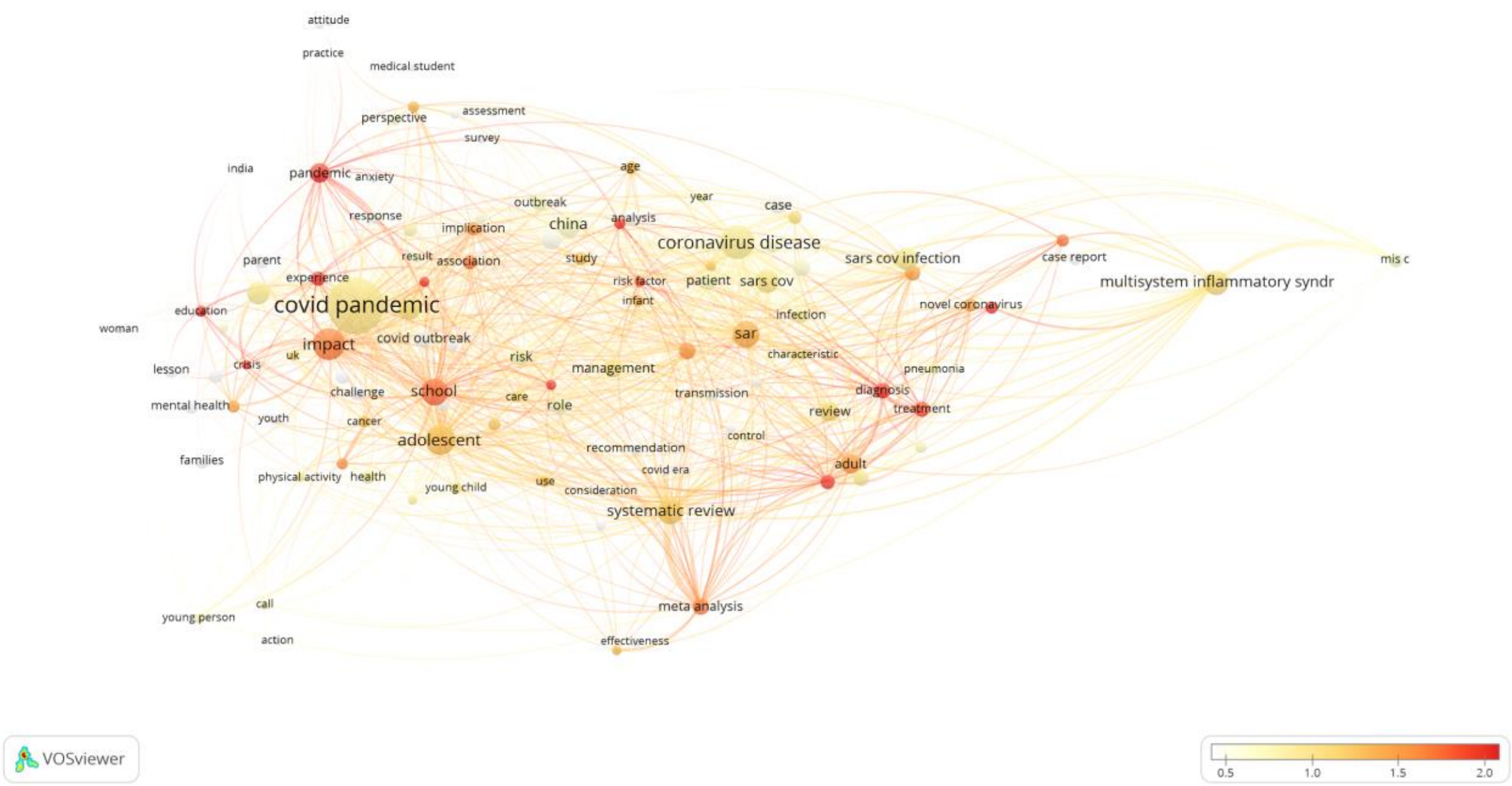

**Figure 4. VOSviewer map of the topics covered by tweets in Spain relative to the overall tweets on children and school closure/reopening. A value of 1 (=yellow nodes) denotes similar focus on the topic as for the overall tweets, nodes in the colors orange-red indicate above average focus. Only top terms are included for visualization purposes (the noun phrases “covid” and “children” were removed since they are redundant).**

The map in Figure 4 depicts more focus than is the case worldwide on topics such as *school* and *education*, but also on *adolescence*, *experience* and *impact*, which suggests a focus on the social implications related to school closure. Furthermore, an enhanced focus on infection and risk factors, diagnosis and treatment is visible on the right-hand side of the overlay map. The severity of symptoms, i.e., *multisystem inflammatory syndrome* appears to be at the same level as for the rest of the world, but at higher levels than for the other two countries. The news overlay map (not shown but uploaded on Figshare) shows less activity around the severity of symptoms and infection, but similar activities for transmission and school, school closure and impact.

## South Africa

Figure 5 provides the timeline for the school closure and (partial) reopenings in South Africa. We note the differentiation between the reopening of secondary (high) schools and primary schools in July and August. Primary schools reopened later than secondary schools. A total of 74 news articles and 992 tweets were identified. Of the 74 news articles, 33 news articles were identified during the first school closure and only 18 in September. The most mentioned studies in the news are those by Yonker et al. (2020) and Hsiang et al. (2020), on large-scale anti-contagion policies, each with 3 mentions. Despite the comparatively low number of identified tweets, we observe Twitter activity around the announcement of school closure in March, and immediately after the first confirmed death, which registers the highest Twitter activity. The highest number of tweets (92 tweets) was in fact in relation to a study on SARS transmission, risk factors and prevention in Hong Kong (Lau et al., 2020), and concerned the wearing of face masks. This illustrates a textbook example on how correlation between two apparent events does not imply causation.

The second peak of Twitter activity, registered in May around the announcement of a phased reopening of schools reveals discussions gravitating either around the Kawasaki-like disease or transmission of the virus by children. Similarly, the discussions from the end of June and beginning of July signal the first Europe-wide study of children (Götzinger et al., 2020) which suggests mild disease in children and very rare fatalities, and relates to the seroepidemiological study in Spain (Pollán et al., 2020). The Spanish study represents the second highest tweeted article, with 39 mentions.

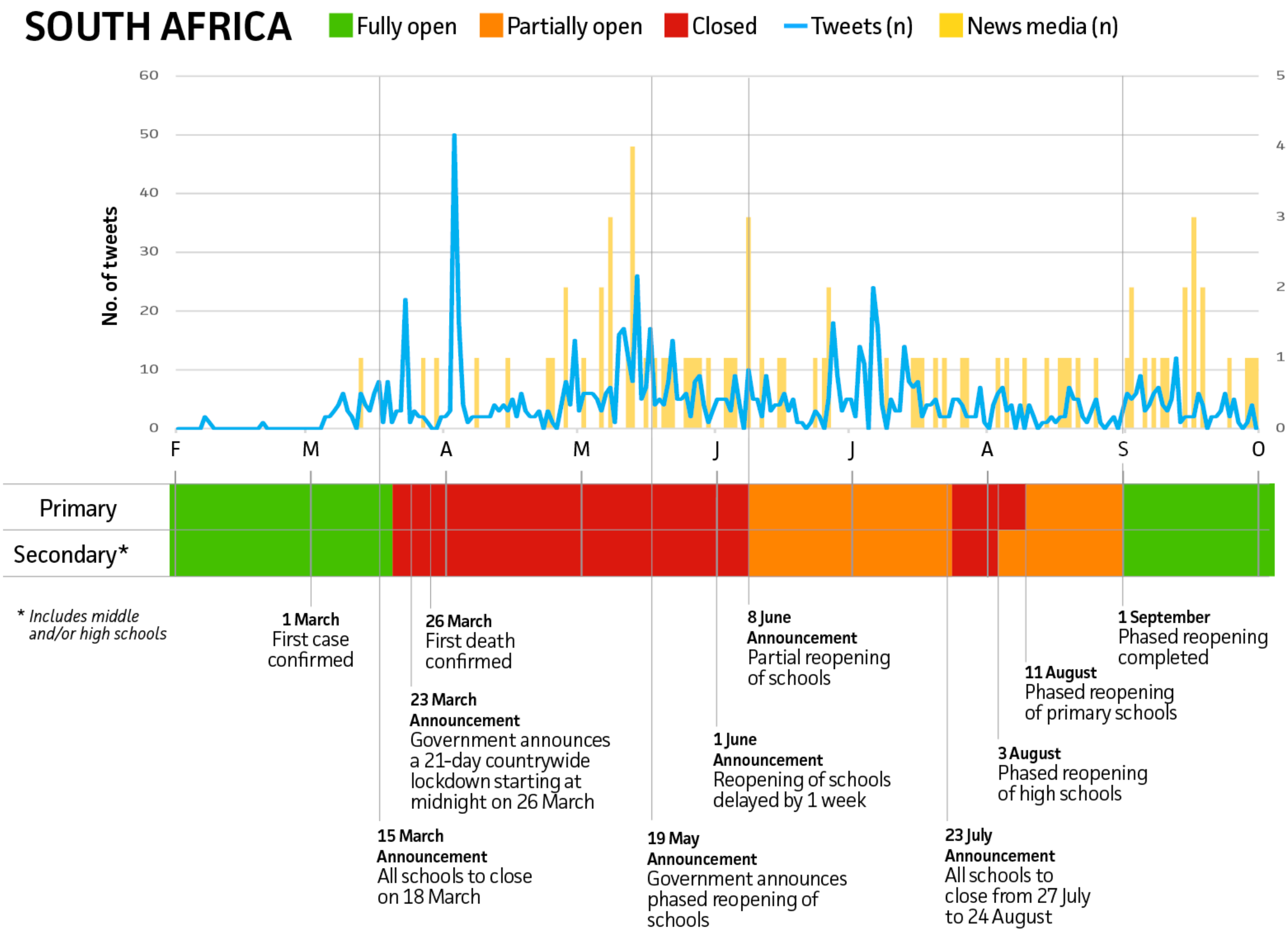

**Figure 5. Timeline of announcements and implementation of school closure and reopening in South Africa, along with the distribution of tweets and news items mentioning scientific articles.**

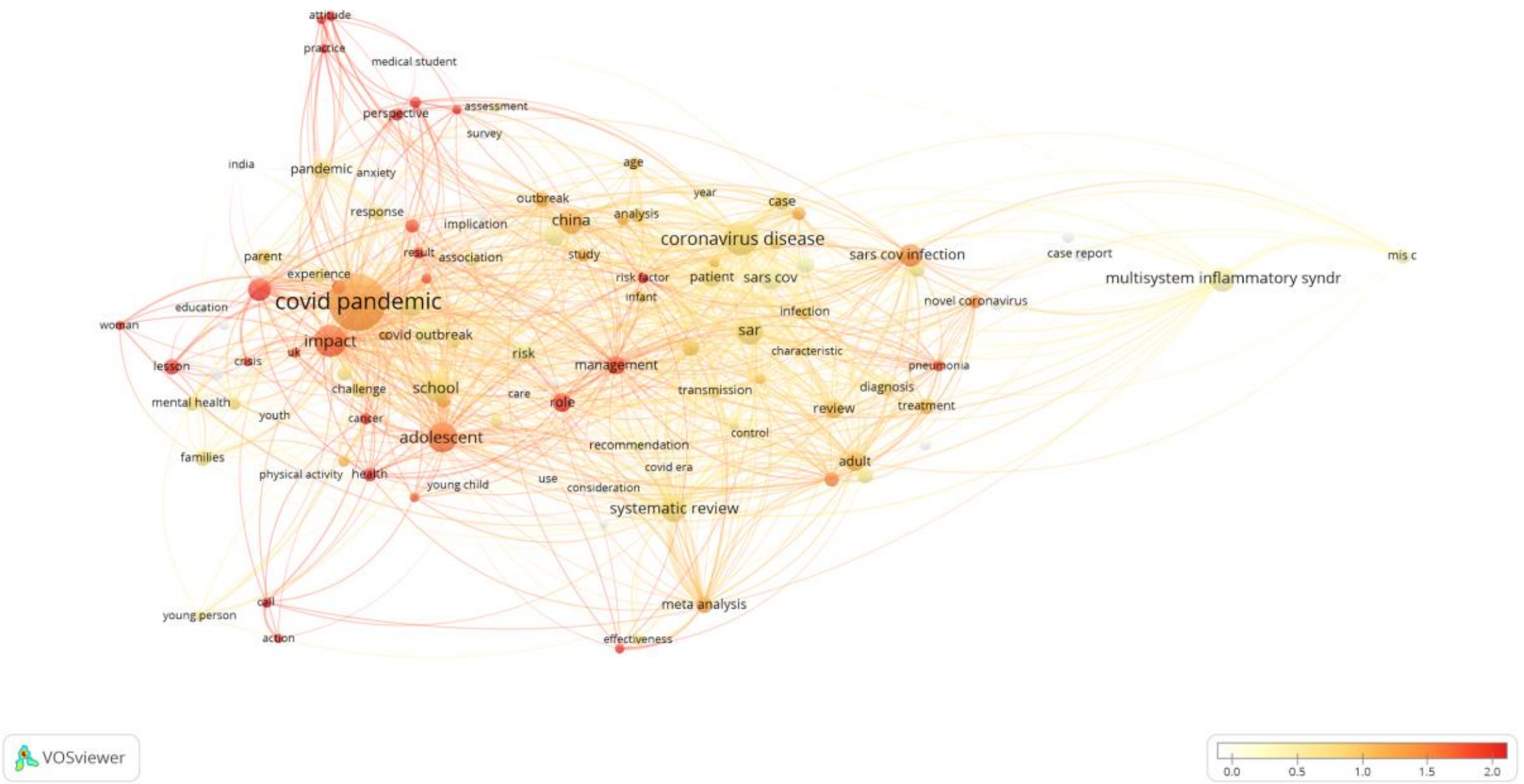

**Figure 6. VOSviewer map of the topics covered by tweets in South Africa relative to the overall tweets on children and school closure/reopening. A value of 1 (=yellow nodes) denotes similar focus on the topic as for the overall tweets, nodes in the colors orange-red indicate above average focus. Only top terms are included for visualization purposes (the noun phrases "covid" and "children" were removed since they are redundant).**

Figure 6 depicts the topics covered by the tweets in South Africa, relative to the total tweets on the topic. Despite the relatively low number of tweets on specific topics, we note that most tweets focused on social aspects (*impact*, *lesson*, *adolescents*), which are depicted on the left-hand side of the map, as well as on *infection*. The focus given to the severity of symptoms was less than in the case of Spain. The news map reflects more attention to the severity of symptoms than captured by the tweet map, similar attention to infection and transmission, as well as to the topic focusing on social aspects, and less attention to 'impact'.

## The Netherlands

Similar to Spain and South Africa, the Netherlands' government decided to close schools in the middle of March. Unlike South Africa, primary schools were reopened first, at half capacity, whereas secondary schools reopened three weeks later, also at half capacity. Additionally, the partial opening was short (on June 8th and June 15th respectively), after which primary and secondary schools were fully open. Finally, it should be noted that the second school closure coincided with the summer holiday in the Netherlands. Figure 7 provides the graphical illustration of the measures, along with the announcements.

The news items and twitter activity over the period is also included. A total of 133 news articles and 1,277 tweets were collected. Most news media attention was dedicated to the topic during the school closure and the partial reopening — 38 news mentions were identified in April and May. Fewer mentions in the news were observed during the summer months, and 21 news mentions were counted in September. The study by Lau et al. (2020) was mentioned 13 times in the news, the highest number of mentions to a scientific article in the Netherlands, supporting the discussion around wearing masks. The authors report that "frequent mask use in public venues", along with other measures have been found to be protective factors. This finding introduces an interesting observation of the discussion around the face masks in the Netherlands, where face masks only became mandatory in December 2020. In Spain and South Africa, where mask wearing was made mandatory earlier on in the pandemic, we do not observe a reaction to this publication, with only one news mention in both Spain and South Africa. A report on coronavirus disease in children from the United States published by the CDC COVID-19 Response Team (Covid CDC et al., 2020) also received 13 mentions in the Dutch news media. The news items referencing this article vary in message. One is titled “Children appear to be less susceptible to corona [virus]”, another reads “The number of children with COVID-19 has risen dramatically over the last five months” and another one “Mounting research paints a bleak picture for schools trying to reopen. Most large schools can expect coronavirus cases within 1 week”.

Netherlands registers a relatively modest Twitter activity similar to South Africa. The first sustained discussion on Twitter was stirred up by the announcement on March 12th that schools will remain open. The tweets debated the government decision and the message transmitted by RIVM that children are less susceptible to become infected. The highest tweet activity was generated by the study by Lau et al. (2020) in early April, with 96 tweets. In the middle of May, a systematic review (Lecrerc et al., 2020) on the available literature on examples of SARS-CoV-2 clusters linked to indoor activities, caught the attention of tweeters in the Netherlands. The majority of 70 tweets focused on the need of outdoor sports (for children), as supported by the lack of empirical evidence for outdoor settings. Finally, the second-most tweeted (79 tweets) article mentioned paediatric SARS-CoV-2 (Yonker et al., 2020) and third-most tweeted output (with 71 tweets) provides scientific support for the wearing of face masks (Peeples, 2020). Even though face masks are not the topic of our study in this chapter, it is interesting to note again that in the Netherlands, face masks were not mandatory at the time of the Twitter discussion.

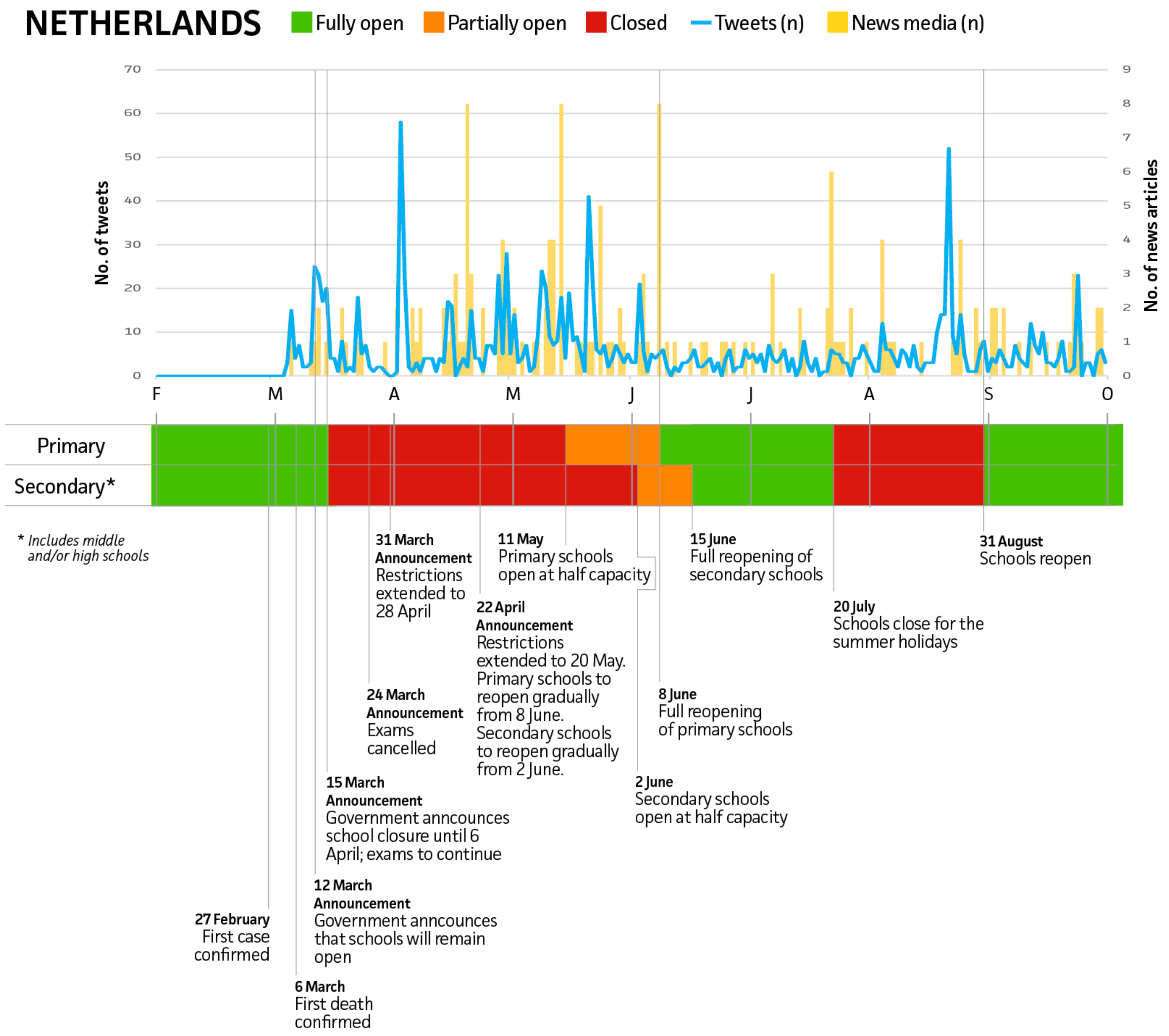

**Figure 7. Timeline of announcements and implementation of school closure and reopening in the Netherlands, along with the distribution of tweets and news items mentioning scientific articles.**

Figure 8 illustrates the topics covered by tweets in the Netherlands relative to the overall tweets on the topic. Unlike in Spain and South Africa, there is little focus on the social topics, which may reflect the early school reopenings in the Netherlands. Most tweets focus on infection and transmission, which are expected topics when schools are open. The news overlay map reflects less activity related to infection and slightly less to severe symptoms, but more to transmission and to school closure than the Twitter overlay map. Similarly, little focus is given to other social aspects.

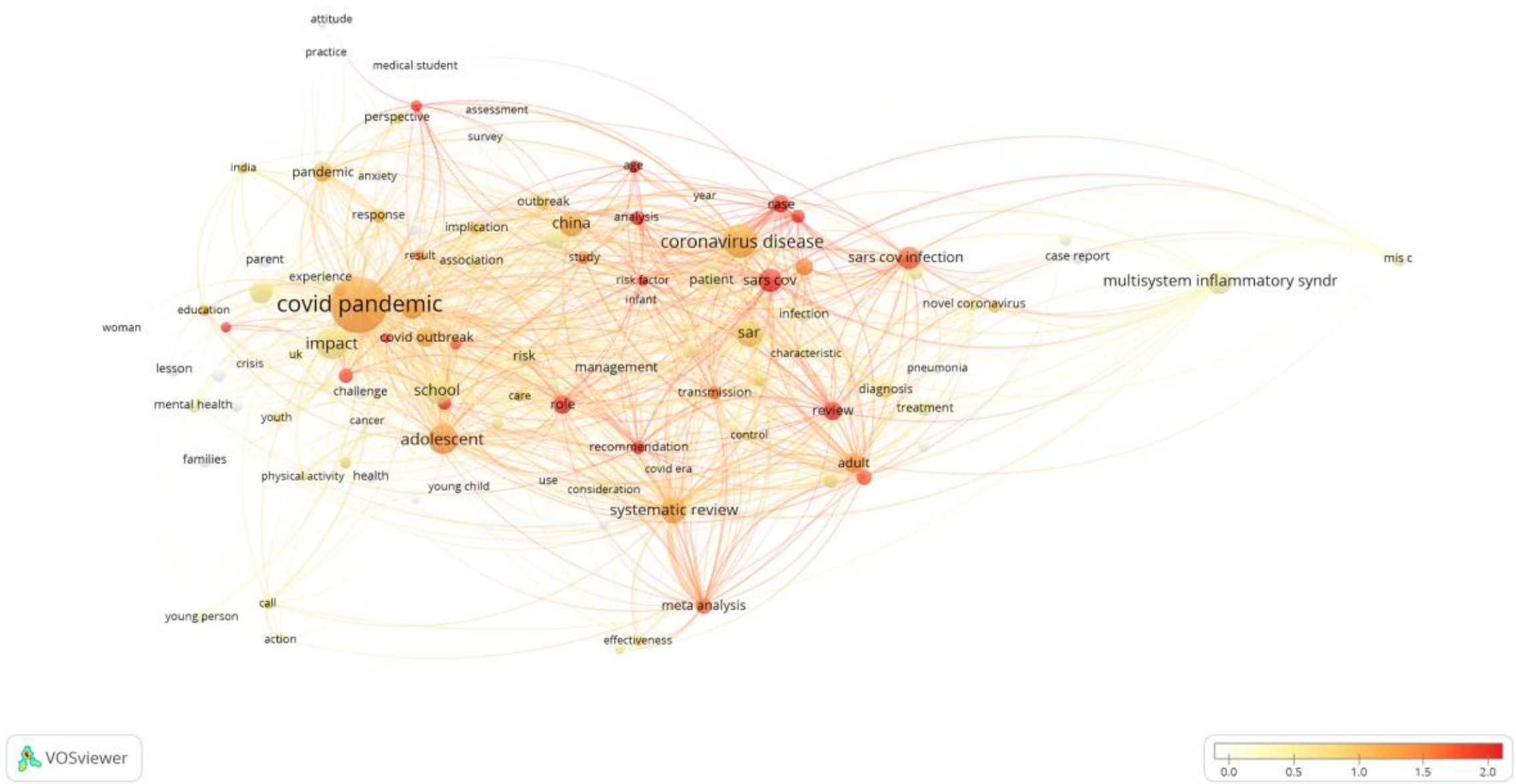

**Figure 8. VOSviewer map of the topics covered by tweets in The Netherlands relative to the overall tweets on children and school closure/reopening. A value of 1 (=yellow nodes) denotes similar focus on the topic as for the overall tweets, nodes in the colors orange-red indicate above average focus. Only top terms are included for visualization purposes (the noun phrases "covid" and "children" were removed since they are redundant).**

## Overall news outlets analysis

The news outlets collected for the three countries were manually scrutinized further. An overall topic analysis based on the titles of the news articles reveals that *infection* appeared in 8 articles in Spain, 12 in the Netherlands, and 2 in South Africa. Kawasaki or other (severe) diseases were mentioned in 7 news articles in the Netherlands, 4 in Spain and 5 in South Africa. *Transmission of the virus* registered the most attention, with 38 news articles in the Netherlands, 29 in Spain and 12 in South Africa. Similarly, *school closur*e was covered by 21 news articles in the Netherlands, 6 in South Africa, and 5 in Spain. This is remarkable given that the Netherlands registered the quickest time to school reopening, which suggests that the news items generally supported the policy measures for schools reopening.

It is also notable that the news items in the three countries do not appear to have covered the scientific output of the national researchers. The only notable exception is the ENE-COVID national study (Pollán et al., 2020), which was mentioned in 20 news articles.

Overall, the most referenced article in the news items of the three countries is the study on paediatric SARS-CoV-2 (Yonker et al., 2020), which is also the second-most tweeted article in the Netherlands. The article was referenced in 25 news articles across all three countries.

Our exploratory analysis of the news articles also revealed cases of miscommunication in reporting the scientific findings in the news media. A first example comes from a recent systematic review and meta-analysis focused on the infection among children compared with adults (Viner et al., 2020a). Their conclusion was that "There is weak evidence that children and adolescents play a lesser role than adults in transmission of SARS-CoV-2 at a population level. This study provides no information on the infectivity of children." A *Nature* news article (Lewis, 2020) reported on the findings as follows: "Researchers suspect that one reason schools have not become COVID-19 hot spots is that children — especially those under the age of 14 — are less susceptible to infection than adults, according to a meta-analysis (Viner et al., 2020a) of prevalence studies."

The study on paediatric SARS-CoV-2 (Yonker et al. 2020) concludes that “children may be a potential source of contagion in the SARS-CoV-2 pandemic despite having milder disease or a lack of symptoms” (p. 45). After finding that regardless of the viral load most children are either asymptomatic or have milder disease, they warn “that it would be ineffective to rely on symptoms or temperature monitoring to identify SARS-CoV-2 infection” (p. 51). While the paper clearly emphasizes the potential risk of children spreading the disease, it does not present evidence of higher transmission rates from children. However, a news item from the Netherlands mentioning the article is entitled “Children with mild symptoms from coronavirus appear to be very infectious”. Four other Dutch news items report the findings of the study, which are not in line with the guidelines from the National Institute for Public Health and Environment (RIVM) that children play a modest role in the transmission of the coronavirus. Spanish media presented a very different perspective on the paper’s findings, warning in their news articles titles that asymptomatic children are spreaders of the disease.

A final example is linked to a Dutch news item entitled “Children ‘do transmit COVID-19’ to adults, says researcher whose study was ‘misunderstood’ as evidence that kids cannot spread coronavirus” published on April 30, 2020. The mentioned study is a review of 78 studies, from which one reports that no child under 10 years of age was found to transmit COVID-19. The article also reports that the German virologist Christian Drosten “pointed out that the findings of the Dutch study from 54 households (RIVM, 2020) which was used as evidence that children do not play a big role in spreading COVID-19, were not statistically significant”.

## Overall Twitter analysis

An overall Twitter analysis on the topics identified in the scientific output reveals further insights. *Infection* has been the most popular topic in the tweets in the three countries, with 3,031 mentions in Spanish tweets, 218 in Dutch and 216 in South African tweets. However, comparatively few tweets mention *community transmission* in relation with children as a vector of transmission. In Spain, we identified 156 mentions, South Africa 41 and in the Netherlands 12 mentions.

By the same token, the transmission of coronavirus among children has received appreciable attention, with 2,170 tweets in Spain, 125 tweets in the Netherlands and 97 in South Africa. In particular, the study by Stein-Zamir et al. (2020), which reports an outbreak in a high school in Israel, was referenced in 714 tweets in Spain — the third-highest tweeted article. The study points out that community transmission is replicated in a school setting and hence the risk of school reopenings while community spread is high. This suggests that despite the strict measures on school reopenings in Spain, the public debate accounted for the risks of less strict measures. The severity of COVID-19 symptoms in children was captured by 550 tweet mentions of *Kawasaki* and *multisystem inflammatory syndrome* in Spain, 47 in the Netherlands and 31 in South Africa. The tweets in the latter countries promoted a recent publication (Viner and Whittaker, 2020).

Munro and Faust’s (2020) paper “Children are not COVID-19 super spreaders: time to go back to school” received appreciable attention in the social media. The article metrics, as observed on November 12, 2020, report that this article had been picked up by 45 news outlets, referenced in 2 policy sources of the Scottish government, and tweeted by 5,430 accounts. In spite of this, the article received modest attention in the three countries in our study: 202 mentions in Spain, 15 mentions in the Netherlands and 3 in South Africa.

We also investigated the profiles of tweeters whose tweets were collected in our dataset. In particular, we looked into the share of tweeters who had tweeted about science (i.e., tweeted an academic publication) before the pandemic, that is, tweeters from our sample present in altmetric data from 2019. We found that 59.3% in Spain, 60.6% in the Netherlands and 65.1% in South Africa had mentioned other scientific articles in their tweets prior to the pandemic. From the total of 8,597 distinct tweeters identified in all three countries, 5,141 had already referenced scientific output before the COVID-19 pandemic. Moreover, we attempted to determine the professions of the 8,597 tweeters.

We identified 740 researchers, 741 health professionals, and 296 journalists based on terms found in the user descriptions of the tweeters. We note a possible overlap between the groups, as someone can be both a health professional and, e.g., hold a PhD (one of the indicators for being a researcher). Given the limited available information on Twitter and the limitation of our search algorithms, we expect that these results are underestimating the true presence of those professions in our dataset.

## Discussion

This study explored the relationship between scientific advances and their societal reception in the context of a global pandemic. We focused on the COVID-19 pandemic and on the specific case of its effects on children. We investigated the publication of scientific findings related to COVID-19 and children, and how this scientific output has been used in news and social media communication following the government measures related to the closure and reopening of schools since the onset of the pandemic until late October 2020. Despite the broad scientific coverage of the topic, both in the news and social media, we observed locally-driven communication in the context of a global pandemic. Our illustrative exploration also identified a disconnect between the policy (political) timeline and the resultant communication in the social media, suggesting that the (social) media reaction to policy moments is not supported by frequent referencing to scientific output on the topic of children and school closure during the COVID-19 pandemic.

Our analysis of the scientific output revealed differences in the science published by researchers in Spain, the Netherlands and South Africa. Despite a global effort to advance knowledge on the COVID-19 virus in order to contain its spread, differences at the national level in terms of the focus of the research published was observed. In other words, our findings suggest that different scientific priorities may emerge during a global pandemic as determined by different socio-economic contexts as in the cases of Spain, the Netherlands and South Africa.

The available scientific output about the role of children and schools in the COVID-19 pandemic has not been picked up in the social media in the three countries in our study to the same degree. We found that only 17.9% of the publications in our database have been tweeted about in the three countries; this is much less than the coverage of about 63.0% of all attention for CORD-19 publications as overall captured by Altmetric (Colavizza et al. 2020). A total of 932 DOIs (16.3% of the scientific output) has been mentioned in the Spanish tweets on the topic. In the Dutch tweets, only 4% of the scientific output (229 articles) has been mentioned, whereas in South Africa 289 articles (5%) have been mentioned.

Furthermore, as the timelines reveal, Twitter activity as it relates to mentions of scientific papers did not mirror notable policy events, apart from weak evidence in the Netherlands. For most cases, where activity was noted, it was triggered by a scientific event (the publication of a paper). This observation tallies with the finding that, based on an analysis of Twitter profiles in the sample, approximately one in five of the Twitter accounts was identified as belonging to a researcher or a health professional. It is conceivable that social media is mirroring science given the number of health, research and other professionals in our sample. Moreover, we were able to identify differences in responses to the scientific output on the topic in the three countries. As in the case of differences in scientific focus between the three countries, a difference in focus in terms of social media content was observed, reinforcing the contextual nature of attention, particularly during a pandemic.

The expectation that scientific papers will be mentioned in the media is not unreasonable given high levels of fear and scientific uncertainty, and since we know that social movements can seek to mobilise action or advocate for a particular ideological position by referring to scientific information in the social media (e.g. anti-vaccination movement) (Van Schalkwyk, 2019). In our study, we found no evidence during the pandemic of a pro- or anti-schooling movement on Twitter. This claim is based on the lack of a universal or popular hashtag related to the topic and which would point to a specific community of attention on Twitter. Further support is provided by the fact that there are no highly active or dominant Twitter accounts to be found in the data. One would expect to find highly active Twitter users in cases where the platform is used to amplify messaging. In the data, only two accounts (both from Spain) were found to have tweeted more than 100 times during the eight-month period. One of the accounts belongs to a paediatrician while the other to the Spanish Society for Paediatric Infectious Diseases. In the absence of an ideologically-motivated group, movement or collective, and some evidence that scientific rather than political activity is the driver of social media activity, Twitter is not in this case being used as a communication platform to amplify messaging about the risks or benefits of children attending school during the COVID-19 pandemic.

A more speculative explanation for the apparent absence of ideologically motivated groups is the demise of Twitter as a platform for politicised science communication in favour of other platforms and/or closed communication channels such as WhatsApp or Facebook groups, especially given the push by Facebook towards private communities as central to the future of the social network (Dwoskin, 2019), and the appeal of private groups to users (Holmes, 2018).

Similar to Twitter, the news media around scientific publications exhibited independent activity from policy moments. Nonetheless, we identified a number of cases of miscommunication of the scientific output in the news media, which illustrate that despite the apparent absence of politically-motivated communities, the intricacies of scientific studies are still not always accurately communicated in the media, and this may lead to distorted messages. These preliminary findings point to the need for further systematic research into the politicization of science communication in specific contexts and across multiple communication platforms during a global pandemic.

## Acknowledgements

The authors would like to thank Altmetric for providing access to data on news and Twitter mentions. Rodrigo Costas is partially funded by the South African DST-NRF Center of Excellence in Scientometrics and Science, Technology, and Innovation Policy (SciSTIP). Jonathan Dudek is partially funded by TU Delft COVID-19 Response Fund.